\documentclass{article} 
\usepackage{iclr2022_conference,times}

\usepackage{amsmath,amsfonts,bm}

\def\eqref#1{equation~\ref{#1}}

\def\1{\bm{1}}

\DeclareMathAlphabet{\mathsfit}{\encodingdefault}{\sfdefault}{m}{sl}
\SetMathAlphabet{\mathsfit}{bold}{\encodingdefault}{\sfdefault}{bx}{n}

\usepackage{hyperref}
\usepackage{url}
\usepackage[ruled, linesnumbered]{algorithm2e}
\usepackage{graphicx}

\usepackage{amsmath,amsbsy,amsfonts,amssymb,amsthm,bm,color}
\newtheorem{theorem}{Theorem}
\newtheorem{remark}{Remark}
\newtheorem{proposition}[theorem]{Proposition}
\newtheorem{lemma}{Lemma}
\newtheorem{assumption}{Assumption}

\theoremstyle{definition}

\newcommand{\lip}{\textrm{Lip}}
\newcommand{\bR}{\mathbb{R}}
\newcommand{\bZ}{\mathbb{Z}}
\newcommand{\bS}{\mathbb{S}}

\newcommand{\dist}{\textrm{dist}}
\newcommand{\be}{\mathbf{e}}

\newcommand{\rev}[1]{\textcolor{blue}{#1}}

\iclrfinalcopy

\title{Provably convergent quasistatic dynamics for mean-field two-player zero-sum games}

\author{Chao Ma, Lexing Ying  \\
Department of Mathematics\\
Stanford University\\
Stanford, CA 94305, USA \\
\texttt{\{chaoma,lexing\}@stanford.edu}
}

\newcommand{\cP}{\mathcal{P}}

\begin{document}

\maketitle

\begin{abstract}
In this paper, we study the problem of finding mixed Nash equilibrium for mean-field two-player zero-sum games. Solving this problem requires optimizing over two probability distributions. We consider a quasistatic Wasserstein gradient flow dynamics in which one probability distribution follows the Wasserstein gradient flow, while the other one is always at the equilibrium. Theoretical analysis are conducted on this dynamics, showing its convergence to the mixed Nash equilibrium under mild conditions. Inspired by the continuous dynamics of probability distributions, we derive a quasistatic Langevin gradient descent method with inner-outer iterations, and test the method on different problems, including training mixture of GANs. 
\end{abstract}

\section{Introduction}
Finding Nash equilibrium has seen many important applications in machine learning, such as generative adversarial networks (GANs)~\citep{goodfellow2014generative} and reinforcement learning~\citep{busoniu2008comprehensive}. In these problems, pure Nash equilibria are usually search for a function $f(x,y)$. Yet, the problems arising from machine learning are usually nonconvex in $x$ and nonconcave in $y$, in which case pure Nash equilibrium may not exist. And even if it exists, there is no guarantee for any optimization algorithm to find it efficiently. This difficulty is reflected in practice, that compared with simple minimization, machine learning applications involving Nash equilibria usually have more complicated behaviors and more subtle dependence on hyper-parameters. For example, stable and efficient training of GANs requires a number of carefully designed tricks~\citep{gao2018deep}.

On the other hand, the mixed Nash equilibrium (MNE) is known to exist in much more general settings, e.g. when the strategy spaces are compact and the payoff function is continuous~\citep{glicksberg1952further}. In the mixed Nash equilibrium problem, instead of taking ``pure strategies'' $x$ and $y$, two ''mixed strategies'' for $x$ and $y$, in the form of probability distributions, are considered, resulting in the following functional,
\begin{equation*}
    \int f(x,y)p(x)q(y)dxdy,
\end{equation*}
where $p$ and $q$ are density functions of probability distributions of $x$ and $y$, respectively. Efforts are invested to develop theoretically endorsed algorithms that can efficiently find MNE for high dimensional problems, with applications on the training of mixture of GANs. In~\citet{hsieh2019finding}, a mirror-descent algorithm is proposed and its convergence is proven. In~\citet{domingo2020mean}, theoretical analysis and empirical experiments are conducted for a gradient descent-ascent flow under a Wasserstein-Fisher-Rao metric and its particle discretization. 

In this paper, we also consider the mixed Nash equilibrium problem, and propose a simple QuasiStatic Wasserstein Gradient Flow (QSWGF) for solving the problem. In our dynamics, we treat $q$ as a component with much faster speed than $p$, hence is always at equilibrium as $p$ moves. With entropy regularization for both $p$ and $q$ (without requirement on the strength of the regularizations), we prove that the QSWGF converges to the unique mixed Nash equilibrium from any initialization (under mild conditions). Furthermore, we show there is a simple way to discretize the QSWGF, regardless of the complexity of the Wasserstein gradient flow of $p$ induced by the fact that $q$ is always at equilibrium. Concretely, a partition function related with $p$ appears in the QSWGF dynamics, and we find an efficient way to approximate the partition function. By discretizing the QSWGF, we derive a particle dynamics with an inner-outer structure, named the QuasiStatic Langevin Gradient Descent algorithm (QSLGD). In QSLGD, after each iteration of the outer problem (for the $x$ particles), the inner loop conducts sufficient iterations to bring the $y$ particles to equilibrium. 
Numerical experiments show the effectiveness of QSLGD on synthetic examples and training mixture of GANs. Our method outperforms the vanilla Langevin gradient descent-ascent method when the entropy regularization is weak. 

As a summary, our two major contributions are:
\begin{enumerate}
\item We propose the quasistatic Wasserstein gradient flow dynamics for mixed Nash equilibrium problems, and show its convergence to the unique Nash equilibrium under weak assumptions. Our result neither requires the entropy regularization to be sufficiently strong, nor assumes the dynamics to converge a priori. 
\vspace{2mm}

\item We derive a simple while practical quasistatic Langevin gradient descent algorithm by discretizing the quasistatic Wasserstein gradient flow, by finding an efficient way to approximate the partition function appearing in the dynamics of $p$. The proposed algorithm is applied on several problems including training mixtures of GANs.
\end{enumerate}

\section{Related work}
The mixed Nash equilibrium problem has a long history, with the proof of its existence dates back to \citet{morgenstern1953theory}. It draws new attention in recent years, especially in the machine learning community, due to the development of GANs~\citep{goodfellow2014generative} and adversarial training~\citep{goodfellow2014explaining}. Training mixture of GANs is already discussed in paper~\citep{goodfellow2014generative}. Some numerical experiments were conducted in~\citep{arora2017generalization}. In~\citet{grnarova2017online}, the authors proposed an online learning approach for training mixture of GANs, and proved its effectiveness for semi-shallow GANs (GANs whose discriminator is a shallow neural network). Yet, rigorous theoretical treatment to an algorithm started from~\citep{hsieh2019finding}, in which a mirror descent method was studied and proven to converge. The implementation of the mirror descent method involves big computational cost that asks for heuristics to alleviate. Later, \citep{domingo2020mean} studied more efficient algorithms under a mixture of Wasserstein and Fisher-Rao metrics. Theoretically, the time average of the dynamics' trajectories is shown to converge to the mixed Nash equilibrium. As a comparison, in this work we show the global convergence of the quasistatic Wasserstein gradient flow without the need of taking time average. Meanwhile, the Wasserstein nature of our dynamics makes it easy to implement as well.  

The Wasserstein gradient flow in the density space has been explored in previous works. For example, \citep{wang2019accelerated} studied the Nesterov’s accelerated gradient flows for probability distributions under the Wasserstein metric, and~\citep{arbel2019kernelized} studied practical implementations of the natural gradient method for the Wasserstein metric. Both works focus on minimization problems instead of min-max problems considered in this work. A more related work is~\cite{lin2021wasserstein}, where a natural gradient based algorithm is proposed for training GANs. Yet, the method still optimizes one generator and one discriminator, searching for pure Nash equilibrium.
Another work that derives algorithms for GANs from a Wasserstein perspective is~\citep{lin2021alternating}.

Another volume of works that studies the Wasserstein gradient flow in the machine learning context is the mean-field analysis of neural networks. This line of works started from two-layer neural networks~\citep{mei2018mean, rotskoff2018neural, chizat2018global, sirignano2020mean}, to deep fully-connected networks~\citep{araujo2019mean, sirignano2021mean, nguyen2019mean, wojtowytsch2020banach}, and residual networks~\citep{lu2020mean, weinan2020machine}. The mean-field formulations treat parameters as probability distributions, and the training dynamics are usually the gradient flow under Wasserstein metric. Attempts to prove convergence of the dynamics to global minima are made~\citep{mei2018mean, chizat2018global, rotskoff2019global}, though in the case without entropy regularization a convergence assumption should usually be made a priori.

\section{The quasistatic dynamics}\label{sec: dyn}
We consider the entropy regularized mixed Nash equilibrium problem, which in our case is equivalent with solving the following minimax problem:
\begin{equation}\label{eqn: problem}
    \min_{p\in\cP(\Omega)}\max_{q\in\cP(\Omega)} \int_{\Omega\times\Omega} K(x,y)p(x)q(y)dxdy + \beta^{-1}\int_\Omega p\log p dx -\beta^{-1}\int_\Omega q\log q dy.
\end{equation}
In (\ref{eqn: problem}), $\Omega$ is a compact Riemannian manifold without boundary, and $\cP(\Omega)$ is the
set of probability distributions on $\Omega$. Since $\Omega$ is compact, any probability
distribution in $\cP(\Omega)$ naturally has finite moments. Let
$E(p,q)=\int_{\Omega\times\Omega}K(x,y)p(dx)q(dy)$, and $S(p)=\int_\Omega p\log p dx$ and
$S(q)=\int_\Omega q\log q dy$ be the (negative) entropy of $p$ and $q$, respectively. Then, the
minimax problem (\ref{eqn: problem}) can be written in short as
\begin{equation}\label{eqn: problem2}
    \min_{p\in\cP(\Omega)}\max_{q\in\cP(\Omega)} E(p,q)+\beta^{-1}S(p)-\beta^{-1}S(q). 
\end{equation}

\begin{remark}
Strictly speaking, in (\ref{eqn: problem}) we should distinguish probability distributions and their
density function (if exist), and the entropy should also be defined using the Radon-Nikodym
derivative with canonical measure. In this paper, since $p$ and $q$ indeed have density functions because of the entropy
regularization, we shall abuse the notation by using $p$ and $q$ to represent both probability
distributions and their density functions.
\end{remark}

The entropy regularizations in (\ref{eqn: problem}) and (\ref{eqn: problem2}) make the problem strongly convex in $p$ and strongly concave in $q$. Hence, there exists a unique Nash equilibrium for the problem. Such results are shown for example by the following theorem from~\citep{domingo2020mean}.
\begin{theorem}\label{thm: NME}(Theorem 4 of~\citep{domingo2020mean})
Assume $\Omega$ is a compact Polish metric space equipped with canonical Borel measure, and that $K$ is a continuous function on $\Omega\times\Omega$. Then, problem (\ref{eqn: problem2}) has a unique Nash equilibrium given by the solution of the following fixed-point problem:
\begin{equation}
p(x) = \frac{1}{Z_p} \exp(-\beta U(x,q)), \quad
q(x) = \frac{1}{Z_q} \exp(\beta V(y,p)), \label{eqn: fixed_point}
\end{equation}
where $Z_p$ and $Z_q$ are normalization constants to make sure $p$ and $q$ are probability distributions, and $U$ and $V$ are defined as 
\begin{equation*}
U(x,q) = \frac{\delta E(p,q)}{\delta p}(x) = \int_\Omega K(x,y)q(y)dy,\quad V(y,p) = \frac{\delta E(p,q)}{\delta q}(y) = \int_\Omega K(x,y)p(x)dx.
\end{equation*}
\end{theorem}

Considering the efficiency in high-dimensional cases, a natural dynamics  of interest to find the Nash equilibrium for (\ref{eqn: problem2}) is the gradient descent-ascent flow under the Wasserstein metric,
\begin{align}
\partial_t p_t &= \nabla\cdot\left(p_t\nabla(U(x,q_t)+\beta^{-1}\log p_t)\right), \nonumber\\
\partial_t q_t &= \nabla\cdot\left(q_t\nabla(-V(y,p_t)+\beta^{-1}\log q_t)\right), \label{eqn: WGAD}
\end{align}
because it can be easily discretized into a Langevin gradient descent-ascent method by treating the PDEs as Fokker-Planck equations of SDEs. 
When $\beta^{-1}$ is sufficiently large, (\ref{eqn: WGAD}) can be proven to converge linearly to the unique MNE of (\ref{eqn: problem2})~\citep{eberle2019quantitative}. However, when $\beta^{-1}$ is small, whether (\ref{eqn: WGAD}) converges remains open. This hinders the application of (\ref{eqn: WGAD}) because in practice the entropy terms are usually used as regularization and are kept small. (We realize that it is proven in~\cite{Domingo2022} when our work is under review.)

In (\ref{eqn: WGAD}), the dynamics of $p$ and $q$ have the same speed. In this work, instead, we
study a quasistatic Wasserstein gradient descent dynamics, which can be understood as a limiting
dynamics when the speed of $q$ becomes faster and faster compared with that of $p$. In this case, at
any time $t$, we assume $q_t$ reaches at the equilibrium of the maximizing problem instantaneously
by fixing $p=p_t$ in (\ref{eqn: problem2}). That is to say, at any time $t$, $q_t$ is determined by
\begin{equation}\label{eqn: q_sub}
    q_t = q[p_t] := \arg\max_{q\in\cP(\Omega)} E(p_t, q) - \beta^{-1}S(q).
\end{equation}
On the other hand, $p_t$ follows the Wasserstein gradient descent flow with $q_t=q[p_t]$ at the equilibrium:
\begin{equation}\label{eqn: p_sub}
    \partial_t p_t = \nabla\cdot\left(p_t \nabla \left(\frac{\delta (E(p_t, q[p_t])-\beta^{-1}S(q[p_t]))}{\delta p_t}+\beta^{-1}\log p_t\right)\right).
\end{equation}
The following theorem shows $q_t=q[p_t]$ can be explicitly written as a Gibbs distribution depending on $p_t$, and thus the free energy in (\ref{eqn: p_sub}) can be simplified to depend on a partition function related with $p_t$.

\begin{theorem}\label{thm: dynamics}
Assume $K$ is continuous on the compact set $\Omega$ and $\beta>0$. Then, for fixed $p_t$ the maximization problem (\ref{eqn: q_sub}) has a unique solution 
\begin{equation}\label{eqn: q_sub2}
    q[p_t](y):= \frac{1}{Z_q(p_t)}\exp(\beta V(y,p_t)),
\end{equation}
where $Z_q(p)$ is a normalization factor, $Z_q(p):=\int \exp(\beta V(y,p))dy$. Moreover, the dynamics (\ref{eqn: p_sub}) for $p_t$ can be written as
\begin{equation}\label{eqn: p_sub2}
    \partial_t p_t = \nabla\cdot\left(p_t\nabla\left(\frac{\delta \beta^{-1}\log Z_q(p_t)}{\delta p_t}+\beta^{-1}\log p_t\right)\right).
\end{equation}
\end{theorem}

Let $F_{p,\beta}(p):=\beta^{-1}\log Z_q(p)+\beta^{-1}S(p)$. By Theorem~\ref{thm: dynamics}, the
dynamics (\ref{eqn: p_sub2}) of $p_t$ is the Wasserstein gradient descent flow for minimizing
$F_{p,\beta}(p)$. By the Proposition~\ref{prop: convex} below, $F_{p,\beta}$ is strongly convex with
respect to $p$. Therefore, it is possible to prove global convergence for the dynamics (\ref{eqn: p_sub2}), and thus the convergence for the quasistatic Wasserstein gradient flow for the minimax problem
(\ref{eqn: problem2}).

\begin{proposition}\label{prop: convex}
For any probability distributions $p_1$, $p_2$ in $\cP(\Omega)$, and any $\lambda\in[0,1]$, we have
\begin{equation*}
    F_{p,\beta}(\lambda p_1+(1-\lambda) p_2) < \lambda F_{p,\beta}(p_1) + (1-\lambda)F_{p,\beta}(p_2).
\end{equation*}
\end{proposition}

In practice the partition function $\log Z_q(p_t)$ in (\ref{eqn: p_sub2}) seems hard to approximate, especially when $\Omega$ is in high dimensional spaces. However, we show in the following proposition that the variation of the partition function with respect to $p_t$ can be written as a simple form involving $q_t$. This property will be used to derive a particle method in Section~\ref{sec: particle}
\begin{proposition}\label{prop: partition}
For any $p\in\cP(\Omega)$, we have
\begin{equation}
    \frac{\delta \beta^{-1}\log Z_q(p)}{\delta p} = U(\cdot, q[p]),
\end{equation}
where $q[p]$ is defined in (\ref{eqn: q_sub2}). Therefore, the dynamics (\ref{eqn: p_sub2}) is equivalent with
\begin{equation}\label{eqn: p_sub3}
    \partial_t p_t = \nabla\cdot\left(p_t\nabla\left(U(x, q[p_t])+\beta^{-1}\log p_t\right)\right).
\end{equation}
\end{proposition}

\section{Convergence analysis}\label{sec: conv}
In this section, we analyze the convergence of the quasistatic dynamics (\ref{eqn: q_sub2}), (\ref{eqn: p_sub2}). First, we make the following assumptions on $K$.
\begin{assumption}\label{assump: K}
Assume $K\in C^\infty(\Omega\times\Omega)$, which means $K$ has continuous derivatives of any order (with respect to both $x$ and $y$).
\end{assumption}
Since $\Omega$ is compact, assumption~\ref{assump: K} implies boundedness and Lipschitz continuity of any derivatives of $K$.

Now, we state our main theorem, which shows the convergence of QSWGF to the Nash equilibrium.
\begin{theorem}\label{thm: main} ({\bf main theorem})
Assume Assumption~\ref{assump: K} holds for $K$. Then, starting from any initial $p_0, q_0\in\cP(\Omega)$, the dynamics (\ref{eqn: q_sub2}), (\ref{eqn: p_sub2}) has a unique solution $(p_t,q_t)_{t\geq0}$, and the solution converges weakly to the unique Nash equilibrium of (\ref{eqn: problem2}), $(p^*,q^*)$, which satisfies the fixed point problem (\ref{eqn: fixed_point}). 
\end{theorem}
Theorem~\ref{thm: main} guarantees convergence of the quasistatic Wasserstein gradient flow for any
$\beta$, giving theoretical endorsement to the discretized algorithm that we will introduce in the
next section. Note that the initialization $q_0$ in the theorem is not important, because we assume
$q$ achieves equilibrium immediately after the initialization.

\begin{remark}
The assumption on $K$'s smoothness can be made weaker. For example, during the proof, up to $4$-th order derivatives of $K$ is enough to give sufficient regularity to the solution of the dynamics.
We make the strong assumption partly to prevent tedious technical analysis so as to focus on the idea and insights.  
\end{remark}

\paragraph{Proof sketch}
We provide some main steps and ideas of the proof of the main theorem in this section. The detailed proof is put in the appendix. 

By the last section, since $q_t$ is always at equilibrium, we only need to considering a Wasserstein gradient descent flow for $F_{p,\beta}(p)$. Therefore, we can build our analysis based on the theories in~\citep{mei2018mean} and~\citep{jordan1998variational}. However, compared with the analysis therein, our theory deals with a new energy term---$\beta^{-1}\log Z_q(p)$, which has not been studied by previous works. From now on, let $E_{p,\beta}(p) = \beta^{-1}\log Z_q(p)$, and $\Psi(\cdot,p)=\frac{\delta E_{p,\beta}(p)}{\delta p}$. By simple calculation we have
\begin{equation}\label{eqn: psi}
    \Psi(x,p) = U(x,q[p]) = \frac{1}{Z_q(p)}\int_{\Omega} K(x,y)\exp\left(\int_{\Omega} \beta K(x,y)p(x)dx\right) dy.
\end{equation}

First, we study the free energy $F_{p,\beta}(p)$, and show that it has a unique minimizer which satisfies a fixed point condition. This is the result of the convexity of $F_{p,\beta}$. We have the following lemma.
\begin{lemma}\label{lm: Fp}
Assume Assumption~\ref{assump: K} holds for $K$. Then, $F_{p,\beta}$ has a unique minimizer $p^*$ that satisfies
$$F_{p,\beta}(p^*) = \inf_{p\in\cP(\Omega)} F_{p,\beta}(p^*).$$
Moreover, $p^*$ is the unique solution of the following fixed point problem,
\begin{equation}\label{eqn: boltzmann}
    p^* = \frac{1}{Z}\exp{\left(-\beta\Psi(x,p^*)\right)},
\end{equation}
where $Z$ is the normalization factor. 
\end{lemma}

Next, we want to show that any trajectory given by dynamics (\ref{eqn: p_sub3}) will converge to the unique minimizer of $F_{p,\beta}$. To achieve this, we first study the existence, uniqueness, and regularity of the solution to (\ref{eqn: p_sub2}), i.e. the trajectory indeed exists and is well behaved. Related results are given by the following lemma.
\begin{lemma}\label{lm: solution}
Assume Assumption~\ref{assump: K} holds for $K$. Then, starting from any initial $p_0\in\cP(\Omega)$, the weak solution $(p_t)_{t\geq0}$ to (\ref{eqn: p_sub2}) exists and is unique. Moreover, $(p_t)$ is smooth on $(0,\infty)\times\Omega$. 
\end{lemma}
The proof of Lemma~\ref{lm: solution} is based on Proposition 5.1 of~\citep{jordan1998variational}. Especially, the existence part is proven using the JKO scheme proposed in~\citep{jordan1998variational}. We consider a sequence of probability distributions given by the following discrete iteration schemes with time step $h$,
\begin{equation*}
p_0^h = p_0,\quad p_{k}^h = \arg\min_{p\in\cP(\Omega)}\left\{\frac{1}{2}W_2^2(p,p_{k-1}^h)+h F_{p,\beta}(p)\right\},\quad k>0,
\end{equation*}
where $W_2(p,q)$ means the 2-Wasserstein distance between probability distributions $p$ and $q$. Let $(p^h_t)_{t\geq0}$ be the piecewise constant interpolations of $(p_k^h)_{k\geq0}$ on time. We show $(p_t^h)$ converges weakly (after taking a subsequence) to a weak solution of (\ref{eqn: p_sub2}) as $h$ tends to $0$. Details are given in the appendix.

Finally, noting that $F_{p,\beta}$ is a Lyapunov function of the dynamics (\ref{eqn: p_sub2}), we have the following lemma showing the convergence of $(p_t)_{t\geq0}$ to the solution of the Boltzmann fixed point problem (\ref{eqn: boltzmann}). This finishes the proof of the main theorem.
\begin{lemma}\label{lm: conv}
Let $(p_t)_{t\geq0}$ be the solution of (\ref{eqn: p_sub2}) from any initial $p_0\in\cP(\Omega)$. Let $p^*$ be the unique minimizer of $F_{p,\beta}$ given by (\ref{eqn: boltzmann}). Then, $p_t$ converges to $p^*$ weakly as $t\rightarrow\infty$. 
\end{lemma}

As a byproduct, since our convergence results does not impose requirement on $\beta$, if one is interested in the minimax problem without entropy regularization,
\begin{equation}\label{eqn: no_entropy}
    \min_{p\in\cP(\Omega)}\max_{q\in\cP(\Omega)} E(p,q),
\end{equation}
then, Theorem 5 in~\citep{domingo2020mean} ensures that the quasistatic dynamics converges to approximate Nash equilibrium of (\ref{eqn: no_entropy}) as long as $\beta^{-1}$ is small enough. Specifically, a pair of probability distributions $(p,q)$ is called $\epsilon$-Nash equilibrium of (\ref{eqn: no_entropy}) if 
\begin{equation*}
    \sup_{q'\in\cP(\Omega)} E(p,q') - \inf_{p'\in\cP(\Omega)} E(p',q)\leq\epsilon.
\end{equation*}
Then, we have the following theorem as a direct results of Theorem 5 in~\citep{domingo2020mean}:

\begin{theorem}\label{thm: approx_ne}
Let $C_K$ be the bound of $K$ that satisfies $|K(x,y)|\leq C_K$ for any $x,y\in\Omega$, and let $\lip(K)$ be the Lipschitz constant of $K$. For any $\epsilon>0$, let $\delta=\epsilon/(2\lip(K))$, and let $V_\delta$ be the volume of a ball with radius $\delta$ in $\Omega$. Then, as long as 
$$\beta> \frac{4}{\epsilon}\log\left(\frac{2(1-V_\delta)}{V_\delta}\left(\frac{4C_K}{\epsilon}-1\right)\right),$$
there exists $T>0$ which depends on $\epsilon$, such that for any $t>T$, the solution $p_t,q_t$ of the dynamics (\ref{eqn: q_sub2}) and (\ref{eqn: p_sub2}) at $t$ satisfies
$$\sup_{q'\in\cP(\Omega)} E(p_t,q') - \inf_{p'\in\cP(\Omega)} E(p',q_t)\leq\epsilon.$$
\end{theorem}

\section{The quasistatic Langevin gradient descent-ascent method}\label{sec: particle}
It is well known that PDEs with the form 
\begin{equation*}
    \partial_t p(t,x) = \nabla\cdot(p(t,x)\mu(t,x)) + \lambda \Delta p(t,x)
\end{equation*}
are Fokker-Planck equations for SDEs
$d X_t = -\mu(t,X_t)dt + \sqrt{2\lambda}dW_t$,
and the solution for the PDE characterizes the law of $X_t$---the solution of the SDE---at any time. This result connects the Wasserstein gradient flow with SDE, and gives a natural particle discretization to approximate the continuous Wasserstein gradient flow. For example, the Wasserstein gradient descent-ascent flow dynamics (\ref{eqn: WGAD}) is the Fokker-Planck equation of the SDEs
\begin{align*}
    dX_t &= -\nabla_x U(X_t,q_t)dt + \sqrt{2\beta^{-1}}dW_t \\
    dY_t &= \nabla_y V(Y_t,p_t)dt + \sqrt{2\beta^{-1}}dW'_t,
\end{align*}
where $p_t$ and $q_t$ are the laws of $X_t$ and $Y_t$, respectively, and $W_t$ and $W'_t$ are two Brownian motions. Note that we have
\begin{equation*}
    \nabla_x U(x,q) = \int_\Omega \nabla_xK(x,y)q(y)dy,\ \ \nabla_y V(y,p) = \int_\Omega \nabla_yK(x,y)p(x)dx.
\end{equation*}
Therefore, i.i.d. picking $X_0^{(i)}\sim p_0$ and $Y_0^{(i)}\sim q_0$ for $i=1,2,...,n$, the particle update scheme, named Langevin Gradient Descent-Ascent (LGDA),
\begin{align}
    X_{k+1}^{(i)} &= X_{k}^{(i)} - \frac{h}{n}\sum\limits_{j=1}^n\nabla_x K(X_k^{(i)}, Y_k^{(j)}) + \sqrt{2h\beta^{-1}}\xi_k^{(i)}, \nonumber\\
    Y_{k+1}^{(i)} &= Y_{k}^{(i)} + \frac{h}{n}\sum\limits_{j=1}^n\nabla_y K(X_k^{(j)}, Y_k^{(i)}) + \sqrt{2h\beta^{-1}}\zeta_k^{(i)}, \label{eqn: particle1}
\end{align}
approximately solves the SDEs, and thus the empirical distributions of $X_k^{(i)}$ and $Y_k^{(i)}$ approximate the solutions of (\ref{eqn: WGAD}) when $n$ is large. Here, $\xi_k^{(i)}$ and $\zeta_k^{(i)}$ are i.i.d. samples from the standard Gaussian. 

\paragraph{Quasistatic Langevin gradient descent method}
Similarly, the dynamics (\ref{eqn: p_sub2}) for $p$ is the Fokker-Planck equation for the SDE
\begin{equation}\label{eqn: sde1}
    dX_t = -\nabla \Psi(x,p_t)dt + \sqrt{2\beta^{-1}}dW_t,
\end{equation}
where $p_t$ is the law of $X_t$. By proposition~\ref{prop: partition} we have $\Psi(x,p_t) = U(x,q[p_t])$.
Hence, (\ref{eqn: sde1}) can be written as 
\begin{equation}\label{eqn: sde2}
    dX_t = -\nabla_x U(X_t, q[p_t]) dt + \sqrt{2\beta^{-1}}dW_t,
\end{equation}
with $q[p_t]$ at the equilibrium of the maximization problem (\ref{eqn: q_sub}), which can be attained by solving the SDE
\begin{equation}\label{eqn: sde_y}
    dY_t = \nabla_y V(Y_t,p_t)dt + \sqrt{2\beta^{-1}}dW'_t
\end{equation}
for sufficiently long time. 
This motivates us to design a quasistatic particle method as a discretization for the quasistatic Wasserstein gradient flow. Specifically, the method consists of an inner loop and an outer loop. The method starts from some particles $X_0^{(i)}$ and $Y_0^{(i)}$, $i=1,2,...,n$, sampled i.i.d. from $p_0$ and $q_0$, respectively. Then, at the $k$-th step, the inner loop conducts enough iterations on the $Y$ particles to solve (\ref{eqn: sde_y}) with $p_t$ fixed (i.e. with the $X$ particles fixed), which drives the empirical distribution of $\{Y_k^{(i)}\}_{i=1}^n$ near equilibrium before each update of the outer loop. Next, the outer loop updates $X_k^{(i)}$ according the SDE (\ref{eqn: sde2}). The algorithm is summarized in Algorithm~\ref{alg: inner_outer}.

\begin{figure}[ht]
\centering
\begin{minipage}{\linewidth}

\IncMargin{1.5em}
\begin{algorithm}[H]
\caption{Quasistatic Langevin gradient descent method (QSLGD)}\label{alg: inner_outer}
\SetAlgoLined
\SetKwInOut{Input}{input}
\SetKwInOut{Output}{output}
\SetKwComment{Comment}{/* }{ */}

\Indm
\Input{$n_x, n_y, k_0, k_1, k_2, T \in \mathbb{N}_+$, $h_x, h_y>0$, $p_0, q_0\in\cP(\Omega)$}
\Output{Final particles $(X_T^{(i)}, Y_T^{(i)})_{i=1}^n$}
\Indp
\BlankLine

Sample $(X_0^{(i)})_{i=1}^{n_x}$ i.i.d. from $p_0$, and $(Y_0^{(i)})_{i=1}^{n_y}$ i.i.d. from $q_0$\;
$Y_{0,0}^{(i)} \gets Y_0^{(i)}$ for $i=1,2,...,n_y$\;
\For{$s\leftarrow 1$ \KwTo $k_0$}{
    $Y_{0,s}^{(i)} \gets Y_{0,s-1}^{(i)} + \frac{h_y}{n_x}\sum\limits_{j=1}^{n_x} \nabla_y K(X_0^{(j)},Y_{0,s-1}^{(i)}) + \sqrt{2h_y\beta^{-1}}\xi$ \Comment*[r]{\bf remark 3.1}
}
$Y_{0}^{(i)} \gets Y_{0,k_0}^{(i)}$ for $i=1,2,...,n_y$\;

\For{$t\leftarrow 1$ \KwTo $T$ }{
    $Y_{t-1,0}^{(i)} \gets Y_{t-1}^{(i)}$ for $i=1,2,...,n_y$\;
    \For{$s\leftarrow 1$ \KwTo $k_1$}{
        $Y_{t-1,s}^{(i)} \gets Y_{t-1,s-1}^{(i)} + \frac{h_y}{n_x}\sum\limits_{j=1}^{n_x} \nabla_y K(X_{t-1}^{(j)},Y_{t-1,s-1}^{(i)}) + \sqrt{2h_y\beta^{-1}}\xi$\;
    }
    \For{$s\leftarrow 1$ \KwTo $k_2$}{
        $Y_{t-1,s+k_1}^{(i)} \gets Y_{t-1,s+k_1-1}^{(i)} + \frac{h_y}{n_x}\sum\limits_{j=1}^{n_x} \nabla_y K(X_{t-1}^{(j)},Y_{t-1,s+k_1-1}^{(i)}) + \sqrt{2h_y\beta^{-1}}\xi$\;
        $\hat{Y}_{t-1}^{((s-1)n_y+i)} \gets Y_{t-1,s+k_1}^{(i)}$, for $i=1,2,...,n_y$\;
    }
    $X_{t}^{(i)} \gets X_{t-1}^{(i)} + \frac{h_x}{k_2n_y}\sum\limits_{j=1}^{k_2n_y} \nabla_y K(X_{t-1}^{(i)},\hat{Y}_{t-1}^{(j)}) + \sqrt{2h_x\beta^{-1}}\xi$ \Comment*[r]{\bf remark 3.2}
    $Y_{t}^{(i)} \gets Y_{t-1, k_1+k_2}^{(i)}$, for $i=1,2,...,n_y$\;
}
\end{algorithm}
\DecMargin{1.5em}
\end{minipage}
\end{figure}

\begin{remark}
Generally speaking, Algorithm~\ref{alg: inner_outer} consists of two nested loops. The inner loop solves $Y$ particles to equilibrium in each step of the outer loop, while the outer loop makes one iteration every time, using the equilibrium $Y$ particles. In the following are some additional explanation for the Algorithm:
\begin{enumerate}
    \item[1]{\bf line 4:}  at the beginning of the algorithm, we conduct $k_0$ additional inner iterations for $Y$, where $k_0$ may be a large number. This is because at the beginning the $Y$ particles are far from equilibrium. In later outer iterations, since each time the $X$ particles only move for a small distance, the $Y$ particles are close to the equilibrium. Therefore, $k_1$ and $k_2$ need not to be large. 
    \item[2]{\bf line 17:} In each inner loop, we conduct $k_1+k_2$ inner iterations for the $Y$ particles, and collect those from the last $k_2$ iterations. We use these $k_2n$ particles in the update of $X$ particles to approximate the distribution $q[p]$. We assume during the last $k_2$ inner iterations the $Y$ particles are at equilibrium. One can take $k_2$ to be $1$ if $n_y$ is large enough, while taking large $k_2$ allows smaller number of $Y$ particles. 
\end{enumerate}
\end{remark}

\subsection{Examples}
In this section, we apply the quasistatic Langevin gradient descent method to several problems.

\begin{figure}
    \centering
    \includegraphics[width=0.32\textwidth, height=0.31\textwidth]{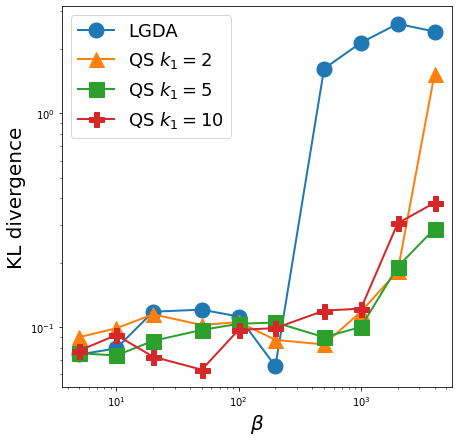}
    \includegraphics[width=0.31\textwidth,height=0.31\textwidth]{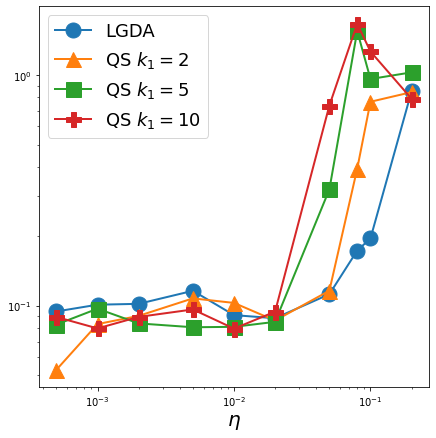}
    \includegraphics[width=0.31\textwidth,height=0.31\textwidth]{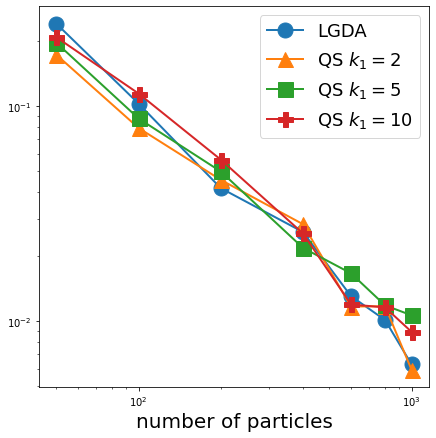}
    \caption{Experiment results with $K(x,y) = \sin(2\pi x)\sin(2\pi y)$. The three figures show the KL divergence of the empirical particle distribution to the uniform distribution of LGDA and QSLGD at different $\beta$, $\eta$ and number of particles. Each point is an average of $5$ experiments.}
    \label{fig: sin}
\end{figure}

\paragraph{$1$-dimensional game on torus}
We first consider a problem with $x$ and $y$ on the $1$-dimensional torus. Specifically, we consider 
\begin{equation*}
    K(x,y) = \sin(2\pi x)\sin(2\pi y),
\end{equation*}
where $x,y\in\bR/\bZ$. It is easy to show that, with this $K$ and a positive $\beta$, at the Nash equilibrium of the problem (\ref{eqn: problem}) $p$ and $q$ are both uniform distributions. We take initial distributions $p_0$ and $q_0$ to be the uniform distribution on $[0,1/4]$. Figure~\ref{fig: sin} shows the comparison of the quasistatic particle method with LGDA for different $\beta$, step length, and number of particles. In the experiments, all quasistatic methods take $k_0=1000$ and $k_2=1$, with different $k_1$ shown in the legends. For each experiment, we conduct $300000$, $150000$, $60000$, $30000$ outer iterations for LGDA, QS2, QS5, and QS10, respectively. We take different different numbers of iterations for different methods in the consideration of different number of inner iterations. The error is then computed after the last iteration, measured by the KL divergence of the empirical distribution given by particles and the uniform distribution (both in the forms of histograms with $10$ equi-length bins). Each point in the figures is an average of $5$ experiments. 

Seen from the left figure, the QSLGD has comparable performance than LGDA when $\beta$ is small, in which case diffusion dominates the dynamics, while it performs much better than LGDA when $\beta$ is large. We can also see better tolerance to large $\beta$ when more inner iterations are conducted. This shows the advantage of the QSLGD over LGDA when the regularization stength is weak. The middle figures shows slightly better performance of the QSLGD when the step length $\eta$ (both $\eta_x$ and $\eta_y$) is small. However, when $\eta$ is big, LGDA tends to give smaller error. The results may be caused by the instability of the inner loop when $\eta$ is big. It also guides us to pick small step length when applying the proposed method. Finally, the right figure compares the influence of the number of particles when $\beta=100$ and $\eta=0.01$, in which case the two methods perform similarly. We can see that the errors for both methods scale in a $1/n$ rate as the number of particles $n$ changes. 

\begin{figure}
    \centering
    \includegraphics[width=0.24\textwidth]{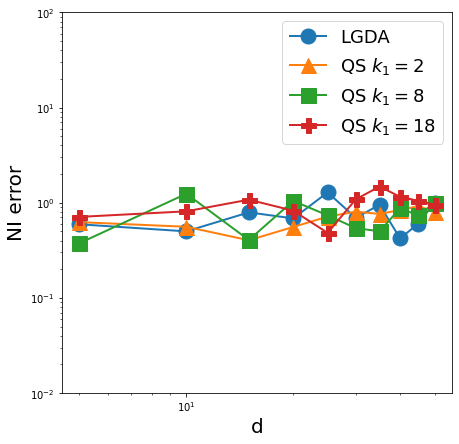}
    \includegraphics[width=0.25\textwidth, height=0.255\textwidth]{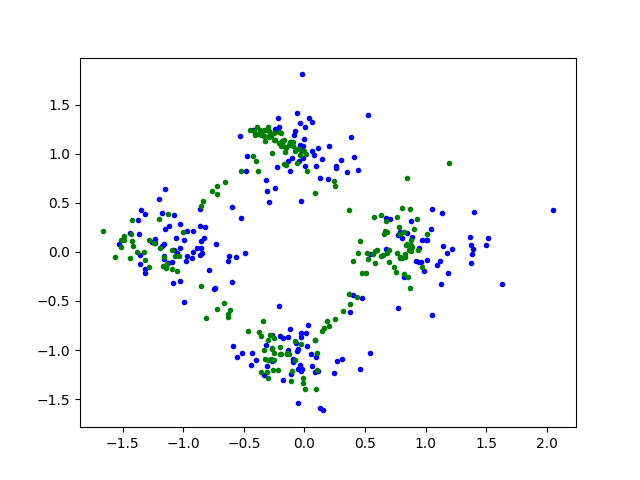}
    \hspace{-3mm}
    \includegraphics[width=0.25\textwidth, height=0.255\textwidth]{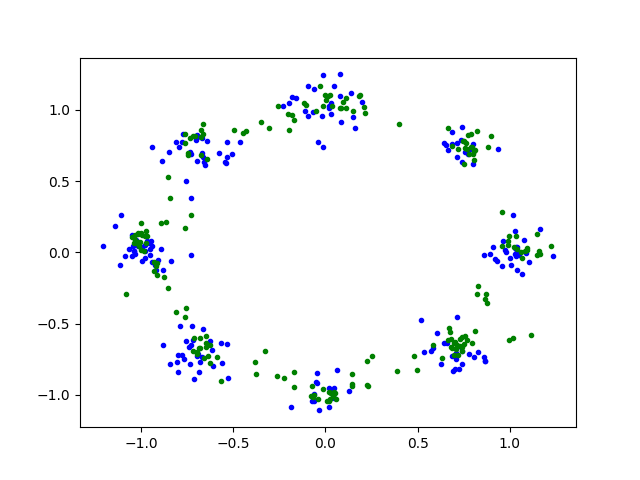}
    \hspace{-2mm}
    \includegraphics[width=0.24\textwidth]{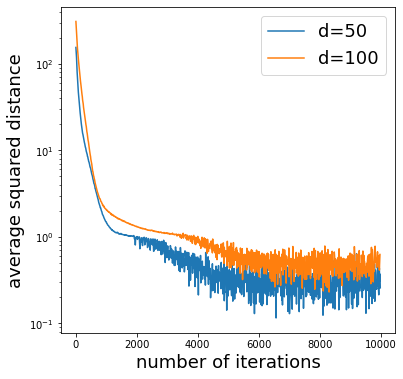}
    \caption{{\bf (Left)} The NI error of the solutions found by different algorithms for the polynomial game (\ref{eqn: poly_game}), at different dimensions. Each point is an average of $10$ experiments. {\bf (Middle left, Middle right)} Generation results of mixture of GANs. The blue points are sampled from groundtruth distribution, while the green points are generated by the a mixture of generators. {\bf (Right)} The average squared distance of generated data to closest mode center for learning high dimensional Gaussian mixtures.}
    \label{fig: poly}
\end{figure}

\paragraph{Polynomial games on spheres}
In the second example, we consider a polynomial games on sphere similar to that studied in~\citep{domingo2020mean},
\begin{equation}\label{eqn: poly_game}
    K(x,y) = x^TA_0x + x^TA_1y + y^TA_2y + y^TA_3(x^2),
\end{equation}
where $x,y\in\bS^{d-1}$ and $(x^2)$ is the element-wise square of $x$. In this problem, we consider the Nash equilibrium of $\min_p\max_q E(p,q)$. Hence, we take big $\beta$ (small $\beta^{-1}$) and compare the Nikaido and Isoda (NI) error of the solutions found by different methods~\citep{nikaido1955note}. The NI error is defined by
\begin{equation*}
    NI(p,q) := \sup_{q'\in\cP(\Omega)} E(p_t,q') - \inf_{p'\in\cP(\Omega)} E(p',q_t),
\end{equation*}
which is also used in Theorem~\ref{thm: approx_ne}. The left panel of Figure~\ref{fig: poly} shows the NI errors of the solutions found by different methods with different dimensions, we see comparable performance of the QSLGD with LGDA.

\paragraph{GANs}
Finally, we test our methods on the training of GANs. We train GANs to learn Gaussian mixtures. The results after training are shown in the middle and right panels of Figure~\ref{fig: poly}, where Gaussian mixtures with $4$ and $8$ modes are learned, respectively. We train GANs with $5$ generators and $5$ discriminators, and take $k_0=100, k_1=5, k_2=1$. The results show that the mixture of GANs trained by QSLGD can learn Gaussian mixtures successfully. 

In the right panel of Figure~\ref{fig: poly}, we show the results of learning high dimensional Gaussian mixtures. In the d-dimensional experiment, the Gaussian mixture has $d$ modes centered at $\be_1, \be_2, ..., \be_d$ with standard deviation $0.1$. Here, $\be_i$ is the $i$-th unit vector in the standard basis of $\bR^d$. Model and algorithm with same hyper-parameters as above are used. In the figure, we measure the average squared distance of the generated data to the closest mode center along the training process. The figure shows that the average squared distance can be reduced to $0.3-0.5$ after $10000$ iterations. While the ideal value is $0.1$, the current results still show that the learnt distribution concentrates at the mode centers. Better results may be obtained after longer training or careful hyper-parameter tuning.

\section{Discussion}
In this paper, we study the quasistatic Wasserstein gradient flow for the mixed Nash equilibrium problem. We theoretically show the convergence of the continuous dynamics to the unique Nash equilibrium. Then, a quasistatic particle method is proposed by discretizing the continuous dynamics. The particle method consists of two nested loops, and conduct sufficient inner loop in each step of the outer loop. Numerical experiments show the effectiveness of the method. Comparison with LGDA shows the proposed method has advantage over LGDA when $\beta$ is large (which is usually the case of interest), and performs as good as LGDA in most other cases. 

Theoretical extensions are possible. For example, strong convergence results may be established by similar approaches taken in~\citep{feng2020entropy}. We leave this as future work.

\rev{In practice,} the idea of nested loops is not new for minimax optimization problems. It is already discussed and utilized in the earliest works for GANs~\citep{goodfellow2014generative}, and Wasserstein GANs~\citep{arjovsky2017wasserstein}. In those works, the discriminator is updated for several steps each time the generator is updated. Our work is different from these works because we consider mixed Nash equilibrium and hence our method is particle based, while their method searches for pure Nash equilibrium. 

Finally, though particle methods finding mixed Nash equilibria have stronger theoretical guarantees, applying these methods to the training of GANs faces the problem of computational cost. With both the generator and discriminator being large neural networks, training mixture of GANs with many generators and discriminators imposes formidable computational cost. Developing more efficient particle methods for GANs is an important future work.

\bibliography{ref}

\begin{thebibliography}{32}
\providecommand{\natexlab}[1]{#1}
\providecommand{\url}[1]{\texttt{#1}}
\expandafter\ifx\csname urlstyle\endcsname\relax
  \providecommand{\doi}[1]{doi: #1}\else
  \providecommand{\doi}{doi: \begingroup \urlstyle{rm}\Url}\fi

\bibitem[Ara{\'u}jo et~al.(2019)Ara{\'u}jo, Oliveira, and
  Yukimura]{araujo2019mean}
Dyego Ara{\'u}jo, Roberto~I Oliveira, and Daniel Yukimura.
\newblock A mean-field limit for certain deep neural networks.
\newblock \emph{arXiv preprint arXiv:1906.00193}, 2019.

\bibitem[Arbel et~al.(2019)Arbel, Gretton, Li, and
  Mont{\'u}far]{arbel2019kernelized}
Michael Arbel, Arthur Gretton, Wuchen Li, and Guido Mont{\'u}far.
\newblock Kernelized wasserstein natural gradient.
\newblock \emph{arXiv preprint arXiv:1910.09652}, 2019.

\bibitem[Arjovsky et~al.(2017)Arjovsky, Chintala, and
  Bottou]{arjovsky2017wasserstein}
Martin Arjovsky, Soumith Chintala, and L{\'e}on Bottou.
\newblock Wasserstein generative adversarial networks.
\newblock In \emph{International conference on machine learning}, pp.\
  214--223. PMLR, 2017.

\bibitem[Arora et~al.(2017)Arora, Ge, Liang, Ma, and
  Zhang]{arora2017generalization}
Sanjeev Arora, Rong Ge, Yingyu Liang, Tengyu Ma, and Yi~Zhang.
\newblock Generalization and equilibrium in generative adversarial nets (gans).
\newblock In \emph{International Conference on Machine Learning}, pp.\
  224--232. PMLR, 2017.

\bibitem[Busoniu et~al.(2008)Busoniu, Babuska, and
  De~Schutter]{busoniu2008comprehensive}
Lucian Busoniu, Robert Babuska, and Bart De~Schutter.
\newblock A comprehensive survey of multiagent reinforcement learning.
\newblock \emph{IEEE Transactions on Systems, Man, and Cybernetics, Part C
  (Applications and Reviews)}, 38\penalty0 (2):\penalty0 156--172, 2008.

\bibitem[Chizat \& Bach(2018)Chizat and Bach]{chizat2018global}
Lenaic Chizat and Francis Bach.
\newblock On the global convergence of gradient descent for over-parameterized
  models using optimal transport.
\newblock \emph{arXiv preprint arXiv:1805.09545}, 2018.

\bibitem[Domingo-Enrich \& Bruna(2022)Domingo-Enrich and Bruna]{Domingo2022}
Carles Domingo-Enrich and Joan Bruna.
\newblock Simultaneous transport evolution for minimax equilibria on measures.
\newblock \emph{arXiv preprint, https://arxiv.org/pdf/2202.06460.pdf}, 2022.

\bibitem[Domingo-Enrich et~al.(2020)Domingo-Enrich, Jelassi, Mensch, Rotskoff,
  and Bruna]{domingo2020mean}
Carles Domingo-Enrich, Samy Jelassi, Arthur Mensch, Grant Rotskoff, and Joan
  Bruna.
\newblock A mean-field analysis of two-player zero-sum games.
\newblock \emph{arXiv preprint arXiv:2002.06277}, 2020.

\bibitem[E et~al.(2020)E, Ma, and Wu]{weinan2020machine}
Weinan E, Chao Ma, and Lei Wu.
\newblock Machine learning from a continuous viewpoint, i.
\newblock \emph{Science China Mathematics}, 63\penalty0 (11):\penalty0
  2233--2266, 2020.

\bibitem[Eberle et~al.(2019)Eberle, Guillin, and
  Zimmer]{eberle2019quantitative}
Andreas Eberle, Arnaud Guillin, and Raphael Zimmer.
\newblock Quantitative harris-type theorems for diffusions and mckean--vlasov
  processes.
\newblock \emph{Transactions of the American Mathematical Society},
  371\penalty0 (10):\penalty0 7135--7173, 2019.

\bibitem[Feng \& Li(2020)Feng and Li]{feng2020entropy}
Qi~Feng and Wuchen Li.
\newblock Entropy dissipation via information gamma calculus: Non-reversible
  stochastic differential equations.
\newblock \emph{arXiv preprint arXiv:2011.08058}, 2020.

\bibitem[Gao et~al.(2018)Gao, Yang, Wang, Sun, Yang, and Zhou]{gao2018deep}
Fei Gao, Yue Yang, Jun Wang, Jinping Sun, Erfu Yang, and Huiyu Zhou.
\newblock A deep convolutional generative adversarial networks (dcgans)-based
  semi-supervised method for object recognition in synthetic aperture radar
  (sar) images.
\newblock \emph{Remote Sensing}, 10\penalty0 (6):\penalty0 846, 2018.

\bibitem[Glicksberg(1952)]{glicksberg1952further}
Irving~L Glicksberg.
\newblock A further generalization of the kakutani fixed point theorem, with
  application to nash equilibrium points.
\newblock \emph{Proceedings of the American Mathematical Society}, 3\penalty0
  (1):\penalty0 170--174, 1952.

\bibitem[Goodfellow et~al.(2014{\natexlab{a}})Goodfellow, Pouget-Abadie, Mirza,
  Xu, Warde-Farley, Ozair, Courville, and Bengio]{goodfellow2014generative}
Ian Goodfellow, Jean Pouget-Abadie, Mehdi Mirza, Bing Xu, David Warde-Farley,
  Sherjil Ozair, Aaron Courville, and Yoshua Bengio.
\newblock Generative adversarial nets.
\newblock \emph{Advances in neural information processing systems}, 27,
  2014{\natexlab{a}}.

\bibitem[Goodfellow et~al.(2014{\natexlab{b}})Goodfellow, Shlens, and
  Szegedy]{goodfellow2014explaining}
Ian~J Goodfellow, Jonathon Shlens, and Christian Szegedy.
\newblock Explaining and harnessing adversarial examples.
\newblock \emph{arXiv preprint arXiv:1412.6572}, 2014{\natexlab{b}}.

\bibitem[Grnarova et~al.(2017)Grnarova, Levy, Lucchi, Hofmann, and
  Krause]{grnarova2017online}
Paulina Grnarova, Kfir~Y Levy, Aurelien Lucchi, Thomas Hofmann, and Andreas
  Krause.
\newblock An online learning approach to generative adversarial networks.
\newblock \emph{arXiv preprint arXiv:1706.03269}, 2017.

\bibitem[Hsieh et~al.(2019)Hsieh, Liu, and Cevher]{hsieh2019finding}
Ya-Ping Hsieh, Chen Liu, and Volkan Cevher.
\newblock Finding mixed nash equilibria of generative adversarial networks.
\newblock In \emph{International Conference on Machine Learning}, pp.\
  2810--2819. PMLR, 2019.

\bibitem[Jordan et~al.(1998)Jordan, Kinderlehrer, and
  Otto]{jordan1998variational}
Richard Jordan, David Kinderlehrer, and Felix Otto.
\newblock The variational formulation of the fokker--planck equation.
\newblock \emph{SIAM journal on mathematical analysis}, 29\penalty0
  (1):\penalty0 1--17, 1998.

\bibitem[Lin et~al.(2021{\natexlab{a}})Lin, Fung, Li, Nurbekyan, and
  Osher]{lin2021alternating}
Alex~Tong Lin, Samy~Wu Fung, Wuchen Li, Levon Nurbekyan, and Stanley~J Osher.
\newblock Alternating the population and control neural networks to solve
  high-dimensional stochastic mean-field games.
\newblock \emph{Proceedings of the National Academy of Sciences}, 118\penalty0
  (31), 2021{\natexlab{a}}.

\bibitem[Lin et~al.(2021{\natexlab{b}})Lin, Li, Osher, and
  Mont{\'u}far]{lin2021wasserstein}
Alex~Tong Lin, Wuchen Li, Stanley Osher, and Guido Mont{\'u}far.
\newblock Wasserstein proximal of gans.
\newblock \emph{arXiv preprint arXiv:2102.06862}, 2021{\natexlab{b}}.

\bibitem[Lu et~al.(2020)Lu, Ma, Lu, Lu, and Ying]{lu2020mean}
Yiping Lu, Chao Ma, Yulong Lu, Jianfeng Lu, and Lexing Ying.
\newblock A mean field analysis of deep resnet and beyond: Towards provably
  optimization via overparameterization from depth.
\newblock In \emph{International Conference on Machine Learning}, pp.\
  6426--6436. PMLR, 2020.

\bibitem[Mei et~al.(2018)Mei, Montanari, and Nguyen]{mei2018mean}
Song Mei, Andrea Montanari, and Phan-Minh Nguyen.
\newblock A mean field view of the landscape of two-layer neural networks.
\newblock \emph{Proceedings of the National Academy of Sciences}, 115\penalty0
  (33):\penalty0 E7665--E7671, 2018.

\bibitem[Mezard \& Montanari(2009)Mezard and Montanari]{mezard2009information}
Marc Mezard and Andrea Montanari.
\newblock \emph{Information, physics, and computation}.
\newblock Oxford University Press, 2009.

\bibitem[Morgenstern \& Von~Neumann(1953)Morgenstern and
  Von~Neumann]{morgenstern1953theory}
Oskar Morgenstern and John Von~Neumann.
\newblock \emph{Theory of games and economic behavior}.
\newblock Princeton university press, 1953.

\bibitem[Nguyen(2019)]{nguyen2019mean}
Phan-Minh Nguyen.
\newblock Mean field limit of the learning dynamics of multilayer neural
  networks.
\newblock \emph{arXiv preprint arXiv:1902.02880}, 2019.

\bibitem[Nikaid{\^o} \& Isoda(1955)Nikaid{\^o} and Isoda]{nikaido1955note}
Hukukane Nikaid{\^o} and Kazuo Isoda.
\newblock Note on non-cooperative convex games.
\newblock \emph{Pacific Journal of Mathematics}, 5\penalty0 (S1):\penalty0
  807--815, 1955.

\bibitem[Rotskoff et~al.(2019)Rotskoff, Jelassi, Bruna, and
  Vanden-Eijnden]{rotskoff2019global}
Grant Rotskoff, Samy Jelassi, Joan Bruna, and Eric Vanden-Eijnden.
\newblock Global convergence of neuron birth-death dynamics.
\newblock \emph{arXiv preprint arXiv:1902.01843}, 2019.

\bibitem[Rotskoff \& Vanden-Eijnden(2018)Rotskoff and
  Vanden-Eijnden]{rotskoff2018neural}
Grant~M Rotskoff and Eric Vanden-Eijnden.
\newblock Neural networks as interacting particle systems: Asymptotic convexity
  of the loss landscape and universal scaling of the approximation error.
\newblock \emph{stat}, 1050:\penalty0 22, 2018.

\bibitem[Sirignano \& Spiliopoulos(2020)Sirignano and
  Spiliopoulos]{sirignano2020mean}
Justin Sirignano and Konstantinos Spiliopoulos.
\newblock Mean field analysis of neural networks: A central limit theorem.
\newblock \emph{Stochastic Processes and their Applications}, 130\penalty0
  (3):\penalty0 1820--1852, 2020.

\bibitem[Sirignano \& Spiliopoulos(2021)Sirignano and
  Spiliopoulos]{sirignano2021mean}
Justin Sirignano and Konstantinos Spiliopoulos.
\newblock Mean field analysis of deep neural networks.
\newblock \emph{Mathematics of Operations Research}, 2021.

\bibitem[Wang \& Li(2019)Wang and Li]{wang2019accelerated}
Yifei Wang and Wuchen Li.
\newblock Accelerated information gradient flow.
\newblock \emph{arXiv preprint arXiv:1909.02102}, 2019.

\bibitem[Wojtowytsch et~al.(2020)]{wojtowytsch2020banach}
Stephan Wojtowytsch et~al.
\newblock On the banach spaces associated with multi-layer relu networks:
  Function representation, approximation theory and gradient descent dynamics.
\newblock \emph{arXiv preprint arXiv:2007.15623}, 2020.

\end{thebibliography}
\bibliographystyle{iclr2022_conference}


\newpage
\appendix
\section{Proofs for Section~\ref{sec: dyn}}
\subsection{Proof of Theorem~\ref{thm: dynamics}}
Note that the free energy in (\ref{eqn: q_sub}) can be written as 
\begin{equation}\label{eqn: free_energy}
    \int_\Omega V(y,p_t)q(y)dy -\beta^{-1}S(q),
\end{equation}
in which the first term is linear with respect to $q$. Hence, a calculation with Lagrange multiplier shows (\ref{eqn: free_energy}) has a unique minimizer $q[p_t]$ with the form of a Gibbs distribution (e.g. see Chapter 4 of~\citep{mezard2009information}):
\begin{equation}
    q[p_t](y)= \frac{1}{Z_q(p_t)}\exp(\beta V(y,p_t)).
\end{equation}

Next, we consider the free energy for $p_t$ when $q$ is at the equilibrium. By (\ref{eqn: free_energy}) we have
\begin{align}
E(p_t,q[p_t]) &= \int_\Omega V(y,p_t)q[p_t](y)dy \nonumber\\
  &= \frac{1}{Z_q(p_t)}\int_\Omega V(y,p_t)\exp(\beta V(y,p_t))dy. \label{eqn: free_energy_pf1}
\end{align}
On the other hand, we have
\begin{align}
\beta^{-1}S(q[p_t]) &= \beta^{-1}\int_\Omega \frac{1}{Z_q(p_t)}\exp(\beta V(y,p_t)) \log\left(\frac{1}{Z_q(p_t)}\exp(\beta V(y,p_t))\right)dy \nonumber\\
  &=\beta^{-1}\int_\Omega \frac{1}{Z_q(p_t)}\exp(\beta V(y,p_t)) \left(\beta V(y,p_t)-\log Z_q(p_t)\right)dy \nonumber\\
  &= \frac{1}{Z_q(p_t)}\int_\Omega V(y,p_t)\exp(\beta V(y,p_t))dy -\beta^{-1}\log Z_q(p_t). \label{eqn: free_energy_pf2}
\end{align}
Combining (\ref{eqn: free_energy_pf1}) and (\ref{eqn: free_energy_pf2}), we obtain
\begin{equation*}
E(p_t,q[p_t])+\beta^{-1}S(p_t)-\beta^{-1}S(q[p_t])=\beta^{-1}\log Z_q(p_t)+\beta^{-1}S(p_t).
\end{equation*}
Therefore, the dynamics of $p_t$ is the Wasserstein gradient descent flow minimizing the free energy $\beta^{-1}\log Z_q(p_t)+\beta^{-1}S(p_t)$, given by
\begin{equation*}
    \partial_t p_t = \nabla\cdot\left(p_t\nabla\left(\frac{\delta \beta^{-1}\log Z_q(p_t)}{\delta p_t}+\beta^{-1}\log p_t\right)\right).
\end{equation*}
This finishes the proof.

\subsection{Proof of Proposition~\ref{prop: convex}}
Since $S(p)$ is strongly convex, it suffices to show $\log Z_q(p)$ is convex. Recall that
\begin{equation*}
    \log Z_q(p) = \log\left(\int_\Omega \exp(\beta V(y,p))dy\right).
\end{equation*}
Note that $V(\cdot,p) = \int K(x,\cdot)p(x)dx$ is linear with respect to $p$, we have
\begin{equation*}
    V(\cdot,\lambda p_1+(1-\lambda)p_2) =\lambda V(\cdot,p_1)+(1-\lambda)V(\cdot,p_2).
\end{equation*}
Hence,
\begin{align}
\log Z_q(\lambda p_1+(1-\lambda)p_2) &= \log\left(\int_\Omega \exp(\beta V(y,\lambda p_1+(1-\lambda)p_2))dy\right) \nonumber\\
  &= \log\left(\int_\Omega \exp(\beta\lambda V(y,p_1))\cdot\exp(\beta(1-\lambda)V(y,p_2))\right) \nonumber\\
  &\leq \log\left(\left(\int_\Omega \exp(\beta V(y,p_1))\right)^{\lambda}\left(\int_\Omega \exp(\beta V(y,p_2))\right)^{1-\lambda}\right) \nonumber\\
  &=\lambda \log Z_q(p_1) + (1-\lambda)\log Z_q(p_2).
\end{align}
The second last line is given by the H\"older inequality.

\subsection{Proof of Proposition~\ref{prop: partition}}
The proposition follows from the following derivations.
\begin{align}
\frac{\delta \beta^{-1}\log Z_q(p)}{\delta p} 
  &= \beta^{-1}\frac{1}{Z_q(p)}\frac{\delta Z_q(p)}{\delta p}\nonumber\\
  &= \beta^{-1}\frac{1}{Z_q(p)}\int_\Omega \exp(\beta V(y,p))\beta K(x,y)dy \nonumber\\
  &= \int_\Omega K(x,y) \frac{\exp(\beta V(y,p))}{Z_q(p)}dy \nonumber\\
  &= \int_\Omega K(x,y)q[p](y)dy \nonumber\\
  &= U(x,q[p]).
\end{align}

\section{Proof of Theorem~\ref{thm: main}}
In this section, we prove our main theorem. The proof will follow the sketch given in Section~\ref{sec: conv}. 
Given the assumptions on $K$ and the conclusions of Lemma~\ref{lm: Fp} and Lemma~\ref{lm: solution}, Lemma~\ref{lm: conv} is a direct result of Lemma 10.12 in~\citep{mei2018mean}, which we will ignore the proof. In the following, we show Lemma~\ref{lm: Fp} and Lemma~\ref{lm: solution}. Some techniques in the proof come from~\citep{jordan1998variational} and~\citep{mei2018mean}.

\subsection{Proof of Lemma~\ref{lm: Fp}}
Our proof follows the proof of Proposition 4.1 in~\citep{jordan1998variational}. First, we show the existence of the minimizer for $F_{p,\beta}$. To see this, note that $K$ is bounded on $\Omega$. Assume $C_K$ is a constant such that $|K(x,y)|\leq C_K$ for any $x,y\in\Omega$. Then, we have
\begin{equation*}
    E_{p,\beta}(p)=\int_\Omega U(x,q[p])p(x)dx = \int_{\Omega\times\Omega} K(x,y)q[p](y)p(x)dxdy\geq -C_K,
\end{equation*}
and 
\begin{equation*}
    S(p)=\int_\Omega p(x)\log p(x)dx\geq \int_\Omega -\frac{1}{e}dx = -\frac{1}{e}.
\end{equation*}
This means $F_{p,\beta}(p)$ is lower bounded, i.e. $\inf_{p}F_{p,\beta}(p)>-\infty$. Hence, we can find a sequence $(p_k)_{k=1}^\infty$ such that 
\begin{equation*}
    \lim_{k\rightarrow\infty} F_{p,\beta}(p_k) = \inf_{p}F_{p,\beta}(p). 
\end{equation*}
Similar to~\citep{jordan1998variational}, we can show boundedness of $\{\int \max\{p_k\log p_k, 0\}dx\}$ and $\{\int p_k^2dx\}$, which implies that $(p_k)$ is uniformly integrable, and thus there exists a weakly convergent subsequence of $(p_k)$. 

Without loss of generality, assume $p_k\rightharpoonup p^*$ in $L^1(\Omega)$. Then we need to show $p^*$ is a minimizer of $F_{p,\beta}$. By~\citep{jordan1998variational}, the entropy term satisfies
\begin{equation*}
    S(p^*)\leq \liminf_{k\rightarrow\infty} S(p_k).
\end{equation*}
Hence, the conclusion follows if $E_{p,\beta}$ is continuous in the weak topology. To show this, first note that for any $p\in\cP(\Omega)$, we have
\begin{equation*}
    \int_\Omega \exp(\beta V(y,p))dy \geq e^{-\beta C_K}.
\end{equation*}
Because the function $\log(x)$ is $1/c$-Lipschitz for $x\in[c,\infty]$, for any $p_k$ we have
\begin{equation*}
    \left|\beta^{-1}\log Z_q(p_k) - \beta^{-1}\log Z_q(p^*)\right|\leq \beta^{-1}e^{\beta C_K}\left|\int_\Omega \left(e^{\beta V(y,p_k)}-e^{\beta V(y,p^*)}\right)dy\right|.
\end{equation*}
By similar boundedness argument, we have
\begin{align*}
\left|\int_\Omega \left(e^{\beta V(y,p_k)}-e^{\beta V(y,p^*)}\right)dy\right| &\leq \int_\Omega \left|e^{\beta V(y,p_k)}-e^{\beta V(y,p^*)}\right|dy \\
  &\leq e^{\beta C_K} \int_\Omega \beta\left|V(y,p_k)-V(y,p^*)\right| dy \\
  &\leq \beta e^{\beta C_K} \int_\Omega \left|\int_\Omega K(x,y)(p_k(x)-p^*(x))dx\right|dy
\end{align*}
Totally we have
\begin{equation*}
 \left|\beta^{-1}\log Z_q(p_k) - \beta^{-1}\log Z_q(p^*)\right|\leq e^{2\beta C_K}\int_\Omega \left|\int_\Omega K(x,y)(p_k(x)-p^*(x))dx\right|dy.
\end{equation*}
Since $K$ is bounded and Lipschitz, it is easy to show that 
\begin{equation*}
    \lim_{k\rightarrow\infty}\int_\Omega \left|\int_\Omega K(x,y)(p_k(x)-p^*(x))dx\right|dy = 0.
\end{equation*}
Therefore, we have 
\begin{equation*}
    F_{p,\beta}(p^*) \leq \liminf_{k\rightarrow\infty}F_{p,\beta}(p_k) = \inf_{p} F_{p,\beta}(p),
\end{equation*}
and thus $F_{p,\beta}(p^*) = \inf_{p} F_{p,\beta}(p)$.

Next, we show $p^*$ satisfies the fixed point condition
\begin{equation}\label{eqn: boltzmann_pf}
    p^* = \frac{1}{Z}\exp{\left(-\beta\Psi(x,p^*)\right)}.
\end{equation}
This follows the proof of Lemma 10.3 in~\citep{mei2018mean}, by first showing $p^*$ has full support on $\Omega$, and then showing
\begin{equation*}
    \Psi(x,p^*)+\beta^{-1}\log p^*(x)
\end{equation*}
is a constant. 

Finally, we show $p^*$ is unique following Lemma 10.4 of~\citep{mei2018mean}. Specifically, we show the Boltzmann fixed point problem (\ref{eqn: boltzmann_pf}) only has one solution by the convexity of $F_{p,\beta}$. Assume (\ref{eqn: boltzmann_pf}) has two different solutions $p_1$ and $p_2$, i.e.
\begin{equation*}
p_1 = \frac{1}{Z(p_1)}\exp{\left(-\beta\Psi(x,p_1)\right)},\ \ p_2 = \frac{1}{Z(p_2)}\exp{\left(-\beta\Psi(x,p_2)\right)}.
\end{equation*}
Then, we have
\begin{align*}
    \log Z(p_1) &= -\beta\Psi(x,p_1)-\log p_1(x),\\
    \log Z(p_2) &= -\beta\Psi(x,p_2)-\log p_2(x).
\end{align*}
Taking difference of the above two equations, and integrating with $p_1-p_2$, we have
\begin{align}
0 &= \int_\Omega (\log Z(p_1)-\log Z(p_2))(p_1(x)-p_2(x))dx \nonumber \\
  &= -\beta\int_\Omega (\Psi(x,p_1)-\Psi(x,p_2))(p_1(x)-p_2(x))dx - \int_\Omega (\log p_1(x)-\log p_2(x))(p_1(x)-p_2(x))dx. \label{eqn: lm1_pf1}
\end{align}
By the monotonicity of $\log x$, the second term of (\ref{eqn: lm1_pf1}) is non-negative, and takes zero only if $p_1=p_2$. For the first term, recall that we have proven in Proposition~\ref{prop: convex} that $E_{p,\beta}$ is convex, we then have
\begin{equation*}
    E_{p,\beta}(p_1)\geq E_{p,\beta}(p_2) + \int_\Omega\Psi(x,p_2)(p_1(x)-p_2(x))dx,
\end{equation*}
and 
\begin{equation*}
    E_{p,\beta}(p_2)\geq E_{p,\beta}(p_1) + \int_\Omega\Psi(x,p_1)(p_2(x)-p_1(x))dx.
\end{equation*}
Taking difference of the two equations gives
\begin{equation*}
    \int_\Omega(\Psi(x,p_1)-\Psi(x,p_2))(p_1(x)-p_2(x))dx \geq0.
\end{equation*}
Therefore, (\ref{eqn: lm1_pf1}) holds if and only if $p_1=p_2$. This finishes the proof of uniqueness of $p^*$, and also completes the proof of Lemma~\ref{lm: Fp}.

\subsection{Proof of Lemma~\ref{lm: solution}}
We show the existence of the weak solution using the JKO scheme used in~\citep{jordan1998variational}. Let $\dist(\cdot,\cdot)$ be the distance metric on $\Omega$. Then, the 2-Wasserstein distance is on $\cP(\Omega)$ is defined as 
\begin{equation*}
    W_2(p_1, p_2) := \left(\inf_{\gamma\in\Gamma(p_1,p_2)}\int_{\Omega\times\Omega} \dist(x_1,x_2)^2d\gamma(x_1,x_2)\right)^{1/2},
\end{equation*}
where $\Gamma(p_1,p_2)$ contains all couplings of $p_1$ and $p_2$, i.e. probability distributions on $\Omega\times\Omega$ with first and second marginals being $p_1$ and $p_2$, respectively. Then, for any $h>0$, we consider the sequence of probability distributions $(p_k^h)_{k=0}^\infty$ obtained by the following iteration scheme:
\begin{equation}\label{eqn: jko}
p_0^h = p_0,\quad p_{k}^h = \arg\min_{p\in\cP(\Omega)}\left\{\frac{1}{2}W_2^2(p,p_{k-1}^h)+h F_{p,\beta}(p)\right\},\quad k>0,
\end{equation}
By similar argument of Proposition 4.1 in~\citep{jordan1998variational}, each minimization problem in (\ref{eqn: jko}) has a unique solution. Hence, $(p_k^h)_{k=0}^\infty$ is uniquely defined. Let $p_t^h$ be the piecewise constant interpolation of $(p_k^h)$ on $t$, i.e.
\begin{equation*}
    p_t^h = p_k^h,\ \ \textrm{for}\ t\in[kh, (k+1)h), 
\end{equation*}
for $k=0,1,2,...$. We now show that there exists a subsequence of $h_n\rightarrow0$ and a $p_t$ such that $p_t^{h_n}\rightharpoonup p_t$ on $(0,T)\times\Omega$ for any $T>0$ and $p_t$ is a weak solution of (\ref{eqn: p_sub2}). This is proven in two steps:
\begin{enumerate}
    \item The existence of weakly convergence subsequence, and
    \item $p_t^h$ approximately satisfies the equation (\ref{eqn: p_sub2}). 
\end{enumerate}

For the first point, we show uniform integrability by showing
\begin{equation}\label{eqn: UI1}
    \int_\Omega \|x\|^2p_k^h(x)dx \leq C
\end{equation}
and 
\begin{equation}\label{eqn: UI2}
    \int_\Omega \max\{p_k^h\log p_k^h, 0\}dx \leq C
\end{equation}
for any $h$ and $k\geq0$, and an absolute constant $C$. Equation (\ref{eqn: UI1}) follows directly from the compactness of $\Omega$. For (\ref{eqn: UI2}), note that for any $h$ and $k\geq0$ we have
\begin{equation*}
    \frac{1}{2}W_2^2(p^h_k, p^h_{k-1})+hF_{p,\beta}(p_k^h) \leq hF_{p,\beta}(p^h_{k-1}),
\end{equation*}
which implies
\begin{equation*}
    F_{p,\beta}(p_k^h)\leq F_{p,\beta}(p_{k-1}^h). 
\end{equation*}
Therefore, 
\begin{align*}
\int_\Omega \max\{p_k^h\log p_k^h, 0\}dx &\leq S(p_k^h) + \int_\Omega \left|\min\{p_k^h\log p_k^h, 0\}\right|dx \\
  &\leq S(p_k^h) + \int_\Omega \frac{1}{e}dx \\
  &\leq F_{p,\beta}(p_k^h) - E_{p,\beta}(p_k^h) + \int_\Omega \frac{1}{e}dx \\
  &\leq F_{p,\beta}(p_0^h) - E_{p,\beta}(p_k^h) + \int_\Omega \frac{1}{e}dx.
\end{align*}
Since $K$ is bounded, $E_{p,\beta}$ is bounded, and thus the above expression is also bounded, which gives (\ref{eqn: UI2}). With (\ref{eqn: UI1}) and (\ref{eqn: UI2}), there exists $p_t(x)$ and a sequence $(h_n)$ with $h_n\rightarrow0$, such that $p_t^{h_n}\rightharpoonup p_t$ in $L^1((0,T)\times\Omega)$ for any $T>0$. Moreover, $p_t\in\cP(\Omega)$ for almost every $T$. By changing $p_t$ on a zero measure set of $t$, we can assume $p_t\in\cP(\Omega)$ for any $t\in(0,\infty)$. With the same analysis as~\citep{jordan1998variational}, the weak convergence can happen for any $t$, i.e. $p_t^{h_n}\rightharpoonup p_t$ in $L^1(\Omega)$ for any $t\in(0,\infty)$.

For the second point, similar to~\citep{jordan1998variational}, consider any vector field $\xi\in C^\infty (\Omega,\Omega)$ and the corresponding flux $\Phi_\tau$ given by
\begin{equation*}
    \partial_\tau \Phi_\tau = \xi(\Phi_\tau),\ \ \Phi_0(x)=x,
\end{equation*}
and let $q_\tau=\Phi_\tau \sharp p_k^h$, then we have
\begin{equation}\label{eqn: lm2_pf2}
    \frac{1}{\tau}\left(\left(\frac{1}{2}W_2^2(p_{k-1}^h, q_\tau)+hF_{p,\beta}(q_\tau)\right)-\left(\frac{1}{2}W_2^2(p_{k-1}^h, p_k^h)+hF_{p,\beta}(p_k^h)\right)\right)\geq0
\end{equation}
for any $\tau>0$. We need to study the limit when $\tau\rightarrow0^+$. By the calculation in~\citep{jordan1998variational} we have
\begin{equation}\label{eqn: lm2_pf3}
    \limsup_{\tau\rightarrow0^+}\frac{1}{\tau}\left(\frac{1}{2}W_2^2(p_{k-1}^h, q_\tau)-\frac{1}{2}W_2^2(p_{k-1}^h, p_k^2)\right) \leq \int_{\Omega\times\Omega} (y-x)\xi(y)d\gamma(x,y),
\end{equation}
and 
\begin{equation}\label{eqn: lm2_pf4}
    \left.\frac{d}{d\tau}S(q_\tau)\right|_{\tau=0}=-\int_\Omega p_k^h \nabla\cdot\xi dx,
\end{equation}
where the $\gamma$ in (\ref{eqn: lm2_pf3}) is the optimal transport between $p_{k-1}^h$ and $p_k^h$. For the $E_{p,\beta}$ term, we have
\begin{align}
\lim\limits_{\tau\rightarrow0^+}\frac{1}{\tau}\left(E_{p,\beta}(q_\tau)-E_{p,\beta}(p_k^h)\right) &= \lim\limits_{\tau\rightarrow0^+}\frac{1}{\beta\tau}\log\left[\frac{\int_\Omega \exp(\beta V(y,q_\tau))dy}{\int_\Omega \exp(\beta V(y,p_k^h))dy}\right] \nonumber\\
  &=\lim\limits_{\tau\rightarrow0^+} \frac{1}{\beta\tau}\left[\frac{\int_\Omega \exp(\beta V(y,q_\tau))dy}{\int_\Omega \exp(\beta V(y,p_k^h))dy}-1\right] \nonumber\\
  &=\frac{1}{\beta Z_q(p_k^h)} \lim\limits_{\tau\rightarrow0^+}\frac{1}{\tau}\left(\int_\Omega (\exp(\beta V(y,q_\tau))-\exp(\beta V(y,p_k^h)))dy \right) \nonumber\\
  &=\frac{1}{\beta Z_q(p_k^h)} \int_\Omega \exp(\beta V(y,p_k^h))\beta\left( \int_\Omega \nabla_xK(x,y)\cdot\xi(x)p_k^h(x)dx \right)dy \nonumber\\
  &=\int_\Omega \nabla_x \Psi(x,p_k^h)\cdot\xi(x)p_k^h(x)dx. \label{eqn: lm2_pf5} 
\end{align}
Combining the above result with (\ref{eqn: lm2_pf3}) and (\ref{eqn: lm2_pf4}), taking both $\xi$ and $-\xi$, we get from (\ref{eqn: lm2_pf2}) that
\begin{equation}\label{eqn: lm2_pf5.1}
\int_{\Omega\times\Omega} (y-x)\xi(y)d\gamma(x,y) + h\int_\Omega \nabla_x \Psi(x,p_k^h)\cdot\xi(x)p_k^h(x)dx-\frac{h}{\beta}\int_\Omega p_k^h \nabla\cdot\xi dx=0
\end{equation}
for any $\xi\in C^\infty(\Omega,\Omega)$. Then, following the derivation in~\citep{jordan1998variational} (proof of Proposition 5.1), as well as the following control
\begin{equation*}
    \sum\limits_{k=1}^N W_2^2(p_{k-1}^h, p_k^h) \leq Ch
\end{equation*}
for any $N$ that satisfies $Nh\leq T$ for some fixed $T$, we can integrate (\ref{eqn: lm2_pf5.1}) over $t$ by viewing $p_k^h$ as $p_t^h$ at appropriate $t$, and take the limit $h_n\rightarrow0$ and show that $p_t$ is a weak solution. During the limit, we need to pay special attention to the second term, i.e. the following limit, which is not dealt with in the reference:
\begin{equation}\label{eqn: lm2_pf6}
    \lim\limits_{n\rightarrow\infty}\int_0^T\int_\Omega \nabla_x\Psi(x,p_t^{h_n})\cdot\xi(x)p_t^{h_n}(x)dx = \int_0^T\int_\Omega \nabla_x\Psi(x,p_t)\cdot\xi(x)p_t(x)dx.
\end{equation}

We prove this by showing $\nabla_x \Psi(x,p_t^{h_n})$ converges to $\nabla_x \Psi(x,p_t)$ uniformly. 

For any fixed $t$, recall that we have
\begin{equation*}
    p_t^{h_n}\rightharpoonup p_t.
\end{equation*}
Therefore, for any $y\in\Omega$, we have
\begin{equation*}
    \int_\Omega K(x,y)p_t^{h_n}(x)dx \rightarrow\int_\Omega K(x,y)p_t(x)dx.
\end{equation*}
Note that $\int_\Omega K(x,y)p(x)dx$ is uniformly continuous with respect to $y$ for any $p\in\cP(\Omega)$, we can conclude that $\int_\Omega K(x,y)p_t^{h_n}(x)dx$ converges to $\int_\Omega K(x,y)p_t(x)dx$ uniformly over $y$. Hence, we have $Z_q(p_t^{h_n})\rightarrow Z_q(p_t)$, and $q[p_t^{h_n}](y)\rightarrow q[p_t](y)$ uniformly. This further implies that
\begin{equation*}
    \nabla_x \Psi(x,p_t^{h_n}) = \int_\Omega \nabla_x K(x,y)q[p_t^{h_n}](y)dy
\end{equation*}
converges to $\nabla_x \Psi(x,p_t)$ for all $x\in\Omega$. This finishes the proof of (\ref{eqn: lm2_pf6}), and also completes the proof that $p_t$ is a weak solution of of equation (\ref{eqn: p_sub2}). 

Now we have proven the existence of the weak solution. The regularity and uniqueness of the solution follows the same analysis of Proposition 5.1 in~\citep{jordan1998variational}.

\end{document}